\documentstyle[multicol,aps,prl,epsf,here]{revtex}
\draft
\begin{document}

\title { {\it Phys.Rev.Lett., accepted}\\
  Supported magnetic nanoclusters: Softlanding of Pd clusters 
         on  a MgO surface}

\author{  M. Moseler$^1$, H. H\"akkinen$^2$ and U. Landman$^2$ \\
       {\it $^1$Theoretical Quantum Dynamics, Faculty of  Physics,
              University of Freiburg,
           D-79104 Freiburg, Germany}\\
           {\it $^2$School of Physics,
             Georgia Institute of Technology, Atlanta, GA 30332-0430}}
\date{\today}
\maketitle
\begin{abstract}
Low-energy deposition of neutral Pd$_{N}$ clusters ($N$$=$$2-7$
and $13$) on a MgO(001) surface F-center (FC) was studied by
 spin-density-functional molecular dynamics simulations. 
The incident clusters are steered by  
 an attractive "funnel"  created by the FC,
resulting in adsorption of the cluster, with
one of its atoms bonded atop of the FC.
The deposited
Pd$_2$-Pd$_6$ clusters
 retain their gas-phase structures, while for N$>$6 
surface-commensurate  isomers are  
energetically more favorable.
Adsorbed clusters with $N > 3$ are found to remain magnetic 
at  the
surface.
\end{abstract}
\begin{multicols}{2}
\narrowtext
Deposition of atomic clusters onto solid surfaces is a 
versatile surface-processing tool, with applications ranging from
"micro-machining" and surface-smoothing to thin-film growth and
fabrication of model
nanocatalysts~\cite{Meiwes,Parent,Heiz,Rattunde,Harbich,Bokwon,Gspann}.
Theoretical investigations, employing most often molecular
dynamics (MD) simulations in conjunction with semi-empirical classical
interatomic potentials, guided many of the above experiments and
provided valuable insights into the microscopic 
mechanisms of the  
 deposition process
\cite{Cleveland,Cheng-Science,Haberland}.
However, in cases where the dominant part of the cluster-surface
interaction involves  surface chemistry (that is, the creation or breaking
of chemical bonds)~\cite{Gspann},  spin-dependent (magnetic)
processes~\cite{Parent}, or  surface defects of electronic
origin (such as a F-center on an ionic surface)~\cite{Heiz,Hak96,Giordano}, a  
full quantum description of the cluster deposition
process is necessary.

Here we report on a first-principles investigation of softlanding of
 Pd$_N$ clusters
(N=2-7 and 13) onto a MgO(001) surface containing a surface
 F-center, FC (oxygen vacancy),  yielding
microscopic details about  the cluster adsorption mechanism and the 
evolution of the atomic,  electronic, and spin degrees of freedom
during the process. 
 We show that the 
 interaction between the Pd cluster and the FC evolves
from an initial  
long-ranged  attractive polarization  into
chemical bonding involving the localized FC electronic state located
in the MgO band gap. A common structural motif for the adsorption
geometry of the smaller clusters ($N$$\leq$$6$) is found where
the cluster retains its gas-phase geometry, whereas the larger
clusters ($N$=7 and 13) adapt upon adsorption to the underlying surface rocksalt
structure. The interaction with the surface
quenches the spin for clusters with $N$$\leq$$ 3$, retains the
gas-phase triplet state (S=1) for $4\leq N\leq 7$, and for $N$$=$$13$ 
the gas-phase nonet (S=4) transforms  to a septet (S=3) state. 
These results are of 
importance  for
 understanding the  activity of Pd/MgO nanocatalysts
\cite{Hakkinen-prl,cyclo}, as well as for future
investigations of supported magnetic clusters and nanoparticles.

The Pd$_N$/MgO system was treated within the framework of
the local-spin-density
functional (LSD) theory, with   norm-conserving
scalar-relativistic pseudo-potentials~\cite{Troullier} and
self-consistent gradient corrections (PBE-GGA)~\cite{Perdew}. For
cluster impact energies below the band gap of the substrate
material, the Born-Oppenheimer (BO) approximation provides  a
faithful description of the collision dynamics and
therefore  we employed the BO-LSD-MD method~\cite{Barnett}  for
the calculation of the electronic structure and the
 nuclear motion 
of the Pd cluster and the substrate.
The  MgO substrate with the F-center was modeled
with a two-layer  {\it ab-initio} cluster Mg$_{13}$O$_{12}$,
embedded  into a lattice of point-charges 
in order to simulate the
long-range Madelung potential\cite{Hakkinen-prl}.
The lattice parameter of the embedding part
of the substrate was fixed to the
experimental value (4.21 \AA) of bulk  MgO.
The Pd$_N$ cluster and the F-center's    4 nearest-neighbor
Mg atoms and 4 nearest-neighbor O atoms of the first layer
were treated dynamically during the deposition. 
To model  the 
heat conductivity  of the extended MgO surface the equations of motion of
the dynamic  surface atoms included an added  damping
term with a damping  constant $\pi \omega_D/6$~\cite{Haberland}, where $\omega_D$ is
the Debye frequency of bulk MgO.

The initial spin states (triplet for $N$$=$2$-$7 and nonet for
$N$$=$$13$) and geometrical structures of the Pd$_N$ clusters were taken from 
 our recent gas-phase study \cite{Moseler-prl}.
The clusters were placed with a random
orientation 4 \AA{} above the FC (measured from the 
cluster atom closest
to the surface)
 and an initial velocity directed perpendicular to the MgO
surface, corresponding to a kinetic energy of
0.1 eV per atom to simulate softlanding conditions.\cite{random}
The spin of the cluster-substrate system was dynamically evaluated
at each MD time step. 
Subsequent to the dynamical evaluation of the deposition process
for about 1 ps the  simulation was stopped, and starting from the last
recorded configuration a corresponding potential 
 energy minimum was located  by
an energy-gradient  optimization with variable spin;
other spin-isomers (SPIs) were optimized
(starting from the aforementioned optimal 
\end{multicols}
\widetext
\begin{figure}
\setlength{\unitlength}{1cm} \epsfxsize=18cm \epsfbox{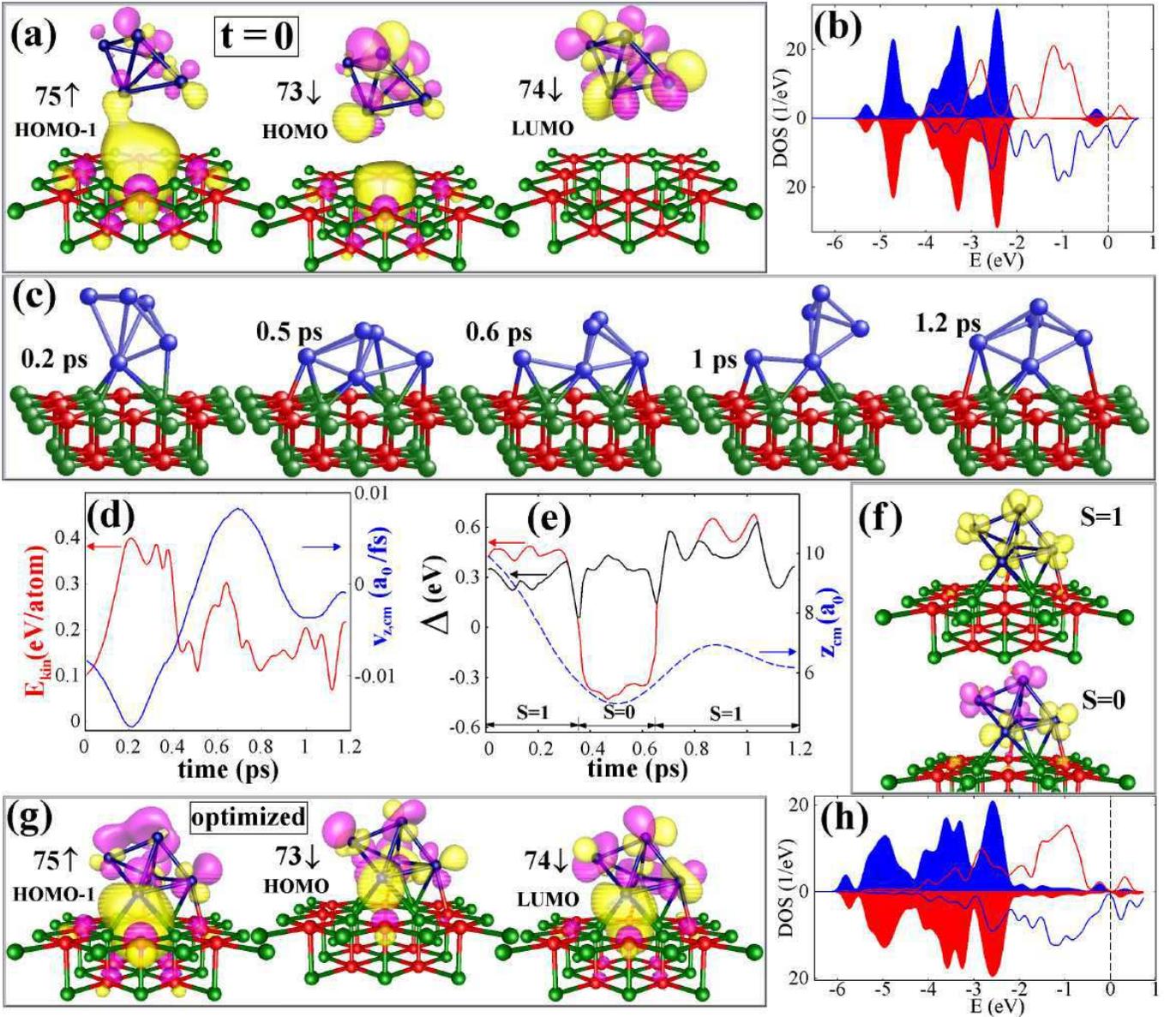}
\caption[]{A Pd$_5$ cluster impinges with 0.1 eV/atom kinetic
energy on an F-center in a MgO (001) surface. Pd atoms are
depicted as blue, Mg as green and O as red spheres. (a)
Isosurfaces of the  highest occupied up-spin molecular orbital (HOMO-1), the  highest
occupied down-spin orbital (HOMO) and the lowest unoccupied down-spin
 orbital (LUMO) of the initial configuration at t$=$0 
(color coding distinguishes the sign of the wave function). Note,
that the 75 up-spin and 73 down-spin orbitals are both occupied.
 (b) The corresponding local densities of states of the surface
(blue area for up- and red area for the down-spin DOS)
and of the
cluster (red line for up- and blue line for down-spin DOS).
The Fermi level, $E_F=0$.
We note that the DOS of the isolated surface and
free cluster are essentially identical to that shown here,
except for the first peak below $E_F$ that corresponds to
the long-range interaction discussed in the text.
 (c) Snapshots from the MD simulation 
recorded at the indicated  times. 
 (d) Time-evolution of the kinetic energy (red line) and the 
$z$-component of the CM  velocity (blue line) of the cluster.
 (e) Evolution of the HOMO-LUMO gap 
(black line), of the eigenvalue energy
 difference $\epsilon_{74\downarrow}-\epsilon_{75\uparrow}$
(red line, see panel (a) for explanation of the
 orbital numbers), and of
 the z-component of the cluster's CM coordinate (blue dashed line).
 The triplet-singlet-triplet
 transition  is indicated by the black arrows
drawn on the time axis.
  (f) Isosurfaces of spin polarization density for the optimized
  triplet (S=1) and singlet (S=0) states.
Yellow and purple denote excess of up and down spins, respectively.
  (g) Isosurfaces of the orbitals in (a) for the optimized adsorbed  cluster.
  (h) The local DOS corresponding to the 
optimized cluster  (color coding as in (b)).
 }
\end{figure}
\begin{multicols}{2}
\narrowtext
configuration),  in order to
explore the thermal stability of the lowest-energy SPI.\\
\indent The adsorption of a single Pd atom on top of the FC
({\it t}FC site) is characterized by a strong 
binding energy (3.31 eV) and a  short equilibrium adsorption distance
(1.65 \AA), compared to adsorption  on-top of an oxygen 
\begin{figure}
\setlength{\unitlength}{1cm} \epsfxsize=8cm \epsfbox{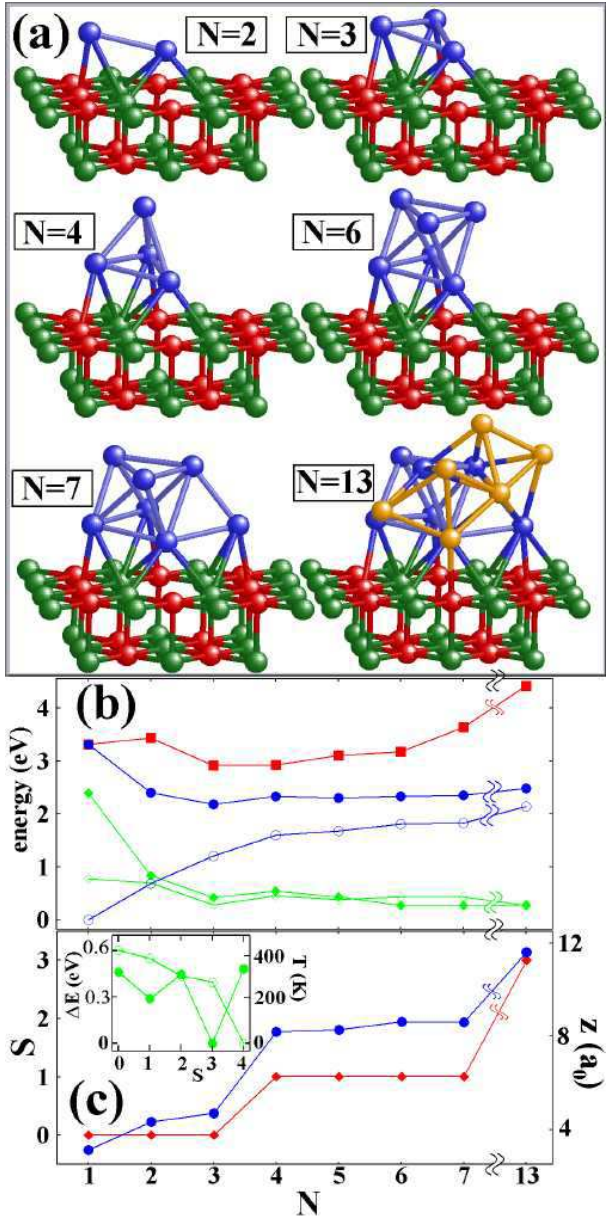}
\caption[]{Structural and magnetic size-evolution of supported
Pd$_N$ clusters. (a) GS structures of Pd$_N$ (N$=$2,3,4,6,7 and
13). Color coding as in Fig.~1 except for Pd$_{13}$ where a subset
of the  Pd atoms is colored in  yellow in order to highlight the
Pd$_7$ subunit (blue).
 (b) Size-evolution of the
adhesion energy $E_{ad}$ (red filled squares), the binding energy
per atom $E_b$ for the supported (blue solid dots)
and free  (blue circles) clusters,  and the HOMO-LUMO gap
of the supported (green solid diamonds) and free (green open
diamonds) clusters.
 (c) Size evolution of GS spin $S$ (red diamonds) and the distance of the
 highest cluster atom to the surface (blue solid dots). The inset in (c) shows
 the SPI energies $\Delta E$ 
(with reference to the GS configuration, $S=3$ for the adsorbed
cluster and $S=4$ for the free one) and corresponding activation temperatures
 $T=2\Delta E/k/(3N-6)$, of supported (green solid dots) and free  Pd$_{13}$
  (green circles) clusters.}
\end{figure}
({\it t}O) atom at
the ideal MgO surface (1.16 eV and 2.17 \AA). The  bonding 
 between the Pd atom and the FC
involves the localized  FC electronic orbital,
located in the
  band gap of MgO (separated from the top of the valence band by
2.3 eV), and (mainly) the $d(m=0)$ orbital of the Pd atom. 
 The attractive interaction to the F-center is rather
long-ranged extending up to about 5 \AA\ above the surface; e.g.,
the interaction energy of a 
Pd atom placed  5.2 \AA\ above the FC 
is   0.1 eV. This weak attraction
 is due to polarization of the
d(m=0)  valence orbital of Pd by the FC. Surprisingly, we found that
none of the other  adsorption sites for the Pd atom, lying in the vicinity
of the F-center (e.g., on-top of the neighboring oxygen ({\it t}O), on-top   of
the neighboring Mg atom ({\it t}Mg), the Mg-Mg bridge ({\it b}MgMg), the
Mg-O bridge ({\it b}MgO) and the Mg-Mg-O hollow site ({\it
h}MgMgO)),  are  stable;
 i.e., optimization starting from any of these
sites leads to a spontaneous (barrierless) transition to
the aforementioned {\it t}FC  configuration. 
We conclude that 
the F-center acts as a rather wide attractive "funnel" for the Pd
atom, extending several \AA\ both laterally and
vertically\cite{FC}. This funneling effect steers 
the incident cluster and  dominates the
dynamics of the  initial
phases of the  deposition process,  as illustrated in the 
following for the representative case  of a   Pd$_5$ cluster.

When the Pd$_5$ cluster is placed 4 \AA\ above the oxygen vacancy, 
the FC electronic state (located just below $E_F$)
combines with d-orbitals of the closest Pd atom to form
two bonding molecular orbitals (see the up-spin
HOMO-1 and the down-spin HOMO in Fig.~1a). All other
orbitals (for example the lowest unoccupied orbital (LUMO) of
Pd$_5$ shown in Fig.~1a) remain to a large degree
 eigenstates  of the separated systems,
and consequently the corresponding density of states (DOS in
Fig.~1b) may be represented as a superposition of those of the bare surface
and the gas-phase cluster.

The  long-range attraction between the cluster and the FC 
 accelerates  the lowermost Pd atom towards the {\it t}FC
site  (note the strong deformation of the  Pd$_5$ cluster  at 0.2 ps in
Fig.~1c and the increase of the kinetic energy in Fig.~1d).
Subsequently, other Pd atoms are attracted to neighboring {\it
b}MgO positions (Fig.~1c, t=0.5 ps) accompanied by additional
release of kinetic energy.
Consequently, the center of mass (CM) velocity toward the surface
increases to almost twice its initial value (Fig.~1d) leading to a
strong flattening of the cluster at t$=$0.5 ps (see the minimum in
the z component of the cluster CM in Fig.~1e). The cluster shape
deformation causes a transient reordering of the molecular 
orbitals, i.e. it raises the energy of the up-spin HOMO-1 level
 (marked 75$\uparrow$
in Fig.~1a)  and turns it into a HOMO at t$=$0.32 ps, and even
into a  LUMO state at t$=$0.36 ps; this sequence is portrayed in 
Fig. 1e by  closing of the HOMO-LUMO
gap (black curve)
and the minimum in the eigenvalue energy difference
$\epsilon_{75\uparrow}$$-$$\epsilon_{74\downarrow}$ 
(red curve). 
Since the down-spin LUMO (74$\downarrow$ in Fig. 1a) temporarily becomes the
HOMO state, the total spin flips for a short time period
 from S=1 to S=0 (Fig.~1e). After 0.5 ps
the cluster recoils  
the reverse process drives the cluster back into the triplet spin
state at t$=$0.63 ps (Fig.~1e).

Optimization of the adsorbed cluster
after a  1.2 ps MD simulation
 (see Fig.~1c for the last MD configuration), 
 resulted in  a
 trigonal bipyramide
 structure (which coincides with the
gas-phase optimal configuration)
  lying with a triangular facet
against a {\it t}FC-{\it b}MgO-{\it b}MgO surface triangle
(Fig.~1f). The spin polarization of the triplet ground state 
(see $S=1$ isosurface in Fig. 1f)
 resembles that of the free cluster with a
minor additional contribution from four  surface oxygen atoms closest to
the FC. As expected from our gas-phase
calculations~\cite{Moseler-prl} the slightly higher-lying singlet
state ($\Delta E$=24 meV) consists of an anti-ferromagnetic ordering
of the local magnetic moments (see $S=0$ isosurface in Fig. 1f). 
The spatial character of the orbitals close to $E_F$ and the 
surface and cluster contributions to the 
DOS of the triplet ground-state of  Pd$_5$/MgO(FC)
are shown in Figs. 1g and 1h, respectively.

Using the above-mentioned methodology, we have determined 
the ground-states for  
 the other deposited
Pd$_N$ clusters~\cite{otherisos}. For  $3\leq N\leq 6$  we
observed a regular structural size evolution (Fig.~2a)  where the
gas-phase GS structures  are anchored to the MgO surface with one Pd
atom on the {\it t}FC, another Pd on 
{\it h}MgMgO site (for $N=2$) or  2 additional Pd
atoms
on {\it b}MgO sites close to the {\it t}O 
position (for $3\leq N\leq 6$). However,
for Pd$_7$ and Pd$_{13}$ the free clusters transform to 
structures that exhibit a higher degree of commensurability
with the underlying surface,
incorporating a Pd$_6$ and Pd$_7$ subunit,
respectively (Fig.~2a).
In this case, the loss in the intracluster cohesion is 
 counterbalanced by a
considerable gain of adhesion energy $E_{ad}$ [defined as
$E_{ad}=E({\rm MgO(FC)})+E({\rm Pd}_N)-E({\rm Pd}_N/{\rm MgO(FC)})$,
 see the red
curve in Fig.~2b]. Consequently, the cohesive energy $E_c$ per Pd
atom [defined as  
$E_c=(E({\rm MgO(FC)})+NE({\rm Pd})-E({\rm Pd}_N/{\rm MgO(FC)}))/N$,
see  the blue curve in Fig. 2b]
continues to  increase after Pd$_6$ and remains well above the
gas-phase $E_c$ values.

The HOMO-LUMO gap (Fig. 2b) of the combined Pd$_N$/MgO(FC) system
is governed  mainly  by  the metal cluster since the top part of its
density of states  lies in the MgO band gap. Most
interestingly, the deposited Pd$_N$ clusters with $N\geq 4$ remain
magnetic: S=1 for $4\leq N \leq 7$ and S=3 for N=13. The crossover
from nonmagnetic to magnetic states between Pd$_3$ and Pd$_4$
correlates with an increased "thickness" of the cluster (Fig.~2c),
corroborating our finding  that flattening of
the cluster on the surface tends to be accompanied by  quenching of the  spin
(see discussion in the context of Fig. 1e).

In general, the deposition of the cluster reduces the energy
separation between SPIs, thus lowering the threshold temperature for  their
coexistence. For instance, the triplet-singlet energy difference
of supported Pd$_4$ is $\Delta E=$65 meV compared to the gas-phase
values of $\Delta E=$136  meV; for Pd$_{13}$ five  SPIs can be found within
a 0.5 eV range, which expressed in terms of temperature 
 corresponds to about 350 K (fig. 2c, inset).
This result indicates that  experiments aiming at
distinguishing magnetic states of the adsorbed clusters could be
carried out at room temperature.

In summary, we have used  first-principles simulations in studies of 
the dynamics of soft-landing of Pd clusters on a MgO(100)
surface containing an F-center defect, 
and  for explorations of the
 structural and spin characteristics of the adsorbed clusters.
 The
F-center creates an attractive "funnel"  for the
approaching metal cluster,
resulting in  preferred binding configurations with
one Pd atom atop  the F-center. A
common structural motif was found for the
smaller  adsorbed clusters (Pd$_2$ - Pd$_6$),
 where the cluster retains its gas-phase geometry.
On the other hand,
Pd$_7$ and Pd$_{13}$ adapt upon adsorption
to the underlying MgO rocksalt
structure. Although the surface tends to reduce
the spin of the adsorbed cluster, clusters larger than Pd$_3$
remain magnetic at the surface, exhibiting several low-lying
structural and spin isomers. These results provide the impetus for
further  
investigations regarding the interplay  of structural and spin
fluxionality of supported metal clusters and their catalytic
properties, and call for an experimental verification of the
predicted magnetic states of  supported Pd clusters.

This research is supported by the U.S. AFOSR and
the U.S. DOE (UL and HH),
the "Deutsche Forschungsgemeinschaft" (MM), and
the Academy of Finland (HH). Simulations were performed at
the NIC (J\"ulich), HLRS (Stuttgart) and NERSC (Berkeley)
computing centers.

%

%
\end{multicols}

\begin{thebibliography}{200}
\bibitem{random}
Since the first-principles simulations are rather
computationally demanding, we have limited ourselves
in most cases to only a few (uncorrelated)
initial starting configurations, characterized by
cluster orientations which are random
with respect to the underlying surface
structure. In all cases the final optimized
configuration of the deposited cluster
 did not depend on the starting
configuration, indicating the dominant
influence of the attractive "funnel"
effect due to the FC, as discussed in the text.
\bibitem{FC} Our study is
relevant to a series
of experiments\cite{Heiz} where nanocatalysts are prepared
by softlanding metal clusters on thin MgO films 
containing typically 
a few \%  ML coverage of FCs. In this case, 
an average
 FC-FC distance is  a few lattice constants.
Therefore, 
 most metal  clusters either experience directly
 the "funnel effect" of
 the nearest FCs while approaching  the  MgO surface, or  become trapped
at  FCs after a  rather short
 surface diffusion path.
\end{thebibliography}
\end{document}